\documentstyle[sprocl]{article}

\def\Journal#1#2#3#4{{#1} {\bf #2}, #3 (#4)}

\def\be{\begin{equation}}
\def\ee{\end{equation}}
\def\bea{\begin{eqnarray}}
\def\eea{\end{eqnarray}}

\let\gam=\gamma

\def\vol{\int d^4x\,\sqrt{-g}} 

\def\half{\frac{1}{2}}

\def\pa{\partial}

\def\PRD{{\em Phys. Rev.} D}
\def\IJMPD{{\em Int. Journ. Mod. Phys.} D}

\def\GRG{\em Gen. Relativ. Grav.}
\def\JPA{\em J. Phys. A: Math. Gen.}
\def\AN{\em Astron. Nach. }
\def\ie{{\it i.e. }}

\def\l{\cal L}
\def\g{\cal G}
\def\fo{\cal F}

\begin{document}

\title{THE EFFECTIVE COSMOLOGICAL CONSTANT IN HIGHER ORDER GRAVITY THEORIES}

\author{S. CAPOZZIELLO$^{1,3}$, R. DE RITIS$^{2,3}$, A.A. MARINO$^{3,4}$}

\address{ $^{1}$Dip. di Scienze Fisiche, "E.R. Caianiello",
 Universit\`{a} di Salerno,\\ I-84081 Baronissi, Salerno,\\  
 $^{2}$Dip. di Scienze Fisiche, Universit\`{a} di Napoli, 
 $^{3}$INFN, Sez.di Napoli,\\
 Mostra d'Oltremare pad. 20 I-80125 Napoli,\\
$^{4}$Oss. Astr. di Capodimonte,
  Via Moiariello, 16 I-80131 Napoli, ITALY\\
E-MAIL: capozziello,deritis,marino@.na.infn.it}
	 
 \maketitle\abstracts{An effective time--dependent cosmological constant can
be recovered for higher--order theories of gravity by extending to these
ones the no--hair conjecture. The results are applied to some specific
cosmological models.}

\noindent 
Fundamental issues as the determination of gravity
vacuum state, the mechanism
which led the early universe to the today observed large scale structures,
and the prediction  of what will be the fate of the whole universe are all 
connected to the determination of cosmological constant.
Besides, from several points of view, it seems that Einstein's theory
of gravity must be generalized to higher--order or scalar--tensor theories
in order to overcome shortcomings of fundamental physics and cosmology.
Both topics have to be connected so that a lot of people is,
recently, asking for the recovering of a sort of cosmological constant
in extended theories of gravity.  With this fact in mind, it is possible
to enlarge the cosmic no--hair conjecture to scalar--tensor~\cite{lambdat}
and higher--order theories by taking into consideration
an "effective" time--dependent cosmological constant which has to 
become asymptotically the "true" one yielding a de Sitter behaviour.
The original no--hair conjecture 
 claims that if there is a positive cosmological 
constant, all
the expanding universes will approach a de Sitter behaviour. 
A simplified version of the conjecture can be
proved ~\cite{wald}. 
It is worthwhile to note that in Wald's paper
 the cosmological constant is a true constant (put by hands)
and the contracted Bianchi 
identity is not used, then the  proof is independent of the
 evolution of  matter. 
In order to extend no--hair conjecture to generalized theories of 
gravity, we have to introduce different sets of conditions
(with respect to those given in~\cite{wald}) since the cosmological 
constant is not
introduced {\it a priori}, but it is "recovered" from  dynamics
of  scalar fields or 
higher--order geometric terms in the gravitational Lagrangians.
 Such conditions must not use
 the "energy conditions", 
but they have to allow the introduction of a sort of effective
cosmological constant which asymptotically becomes the de Sitter constant.
Furthermore, it is the only term that does not 
decrease with time.
Hence, in an expanding universe, $\Lambda$ is the asymptotically dominant term
in the Einstein equations.
Then, given any extended theory of gravity, it could be
possible, in general, to define an effective time varying
 cosmological constant which becomes the "true" cosmological constant
if and only if the model asymptotically approaches de Sitter
(that is only asymptotically no--hair conjecture is recovered). This
fact will introduce constraints on the choice of the gravitational couplings,
scalar field potentials and higher--order geometrical terms which combinations
can be intended as components of the effective stress--energy tensor.   
Here we  deal with higher--order theories asking for 
recovering  the de Sitter behaviour in the related cosmological models,
using the scheme which we adopt   
for scalar--tensor gravity ~\cite{lambdat}.
We take into account the  function $f(R)$ where $R$ is the Ricci scalar.
Let us start from the action
\be 
\label{3.7.1}
{\cal A}=\vol \left[f(R)+{\l}_{m}\right]\,.
\ee 
The standard (minimally coupled) matter 
gives
no contribution to  dymnamics when we consider the asymptotic behaviour
of  system. 
For the sake of simplicity, we discard its
contribution (\ie ${\l}_{m}=0$) from now on.
We adopt a FRW metric considering that the results can be
 extended to any Bianchi model~\cite{lambdat}. 
The Lagrangian density in
action (\ref{3.7.1}) can be written as
${\l}= {\l}(a,\dot{a},R,\dot{R})$,
where $a$ and $R$ are canonical variables. 
Such a  position seems
arbitrary, since $R$ is not independent of $a$ and $\dot{a}$,
but it is generally used in canonical quantization of higher--order 
gravitational theories~\cite{vilenkin}.
In practice, the definition of $R$ by $\ddot{a},\dot{a}$ and $a$
introduces a constraint which
 eliminates the second and higher order derivatives in time, 
then this last one produces
a system of second order differential equations in
$\{a,R\}$.
In fact, using the Lagrange multiplier $\lambda$, we have
that the action can be written as
\be 
 \label{3.7.4}
{\cal A}=2\pi^{2}\int dt\left\{f(R)a^{3}-\lambda\left[
R+6\left(\frac{\ddot{a}}{a}+\frac{\dot{a}^{2}}{a^{2}}+\frac{k}{a^{2}}\right)
\right]\right\}\,.
\ee 
The prime indicates  the derivative with respect to $R$.
In order to determine $\lambda$, we have to vary the action with respect 
to $R$, from which $\lambda =a^{3}f'(R)$.
Substituting into (\ref{3.7.4}) and integrating by parts, we obtain the
Lagrangian
\be 
 \label{3.7.7}
{\l}=a^{3}\left[f(R)-Rf'(R)\right]+6\dot{a}^{2}af'(R)+
6a^{2}\dot{a}\dot{R}f''(R)-6akf'(R)\,.
\ee 
Then the equations of motion are
\be 
\label{3.7.7'}
\left(\frac{\ddot{a}}{a}\right)f(R)'+
2\left(\frac{\dot{a}}{a}\right)f(R)''\dot{R}+f(R)''\ddot{R}+f(R)'''\dot{R}^{2}-
\frac{1}{2}[Rf(R)'+f(R)]=0\,,
\ee 
\be 
 \label{3.7.8}
R=-6\left(\frac{\ddot{a}}{a}+\frac{\dot{a}^{2}}{a^{2}}+\frac{k}{a^{2}}\right)\,,
\ee 
which is the constraint, and
\be 
 \label{3.7.9}
6\dot{a}^{2}af'(R)-a^{3}\left[f(R)-Rf'(R)\right]+
6a^{2}\dot{a}\dot{R}f''(R)+akf'(R)=0\,.
\ee 
Let us now define the auxiliary field $p\equiv f'(R)$, 
so that the Lagrangian (\ref{3.7.7}) can be recast in the form
\be 
 \label{3.7.10}
{\l}=6a\dot{a}^{2}p+6a^2\dot{a}\dot{p}-6akp-a^3W(p)\,,
\ee 
where 
\be 
 \label{3.7.11}
W(p)= h(p)p-r(p)\,,\;\;\;
r(p)=\int f'(R)dR=\int p dR=f(R)\,,\;\;\;
h(p)=R\,,
\ee 
such that $h=(f')^{-1}$ is the inverse function of $f'$.
A Lagrangian like (\ref{3.7.10}) is a special kind of the so called 
Helmholtz Lagrangian~\cite{magnano}.
Dynamical system  becomes
\be 
 \label{3.7.14}
6\left[\dot{H}+2H^{2}
\right]=-\frac{d W(p)}{dp}\,,
\ee 
\be 
 \label{3.7.17}
H^2+H\frac{\dot{p}}{p}+\frac{W(p)}{6p}=0\,,
\ee 
\be 
 \label{3.7.20}
\dot{H}=-\frac{1}{2}\left(H^2+\frac{W(p)}{6p}\right)-
\frac{1}{2}\left(\frac{\dot{p}}{p}\right)^2 -
\half\frac{d}{dt}\left(\frac{\dot{p}}{p}\right)\,.
\ee 
Eq.(\ref{3.7.14}) has the role of the Klein--Gordon equation.
$H$ is the Hubble parameter.
We want to obtain an effective cosmological constant.
For semplicity, we have assumed $k=0$. Eq.(\ref{3.7.17}) 
 can be recast as
\be 
 \label{3.7.18}
(H-\Lambda_{eff,\,1})(H-\Lambda_{eff,\,2})=0\,.
\ee 
The effective cosmological constant can be formally defined as
\be 
 \label{3.7.19}
\Lambda_{eff\, 1,2}=-\frac{\dot{p}}{2p}
\pm\sqrt{\left(\frac{\dot{p}}{2p}\right)^2-\frac{W(p)}{6p}}\,.
\ee 
We have to note that Eq.(\ref{3.7.18}) defines the exact solutions
$H(t)=\Lambda_{eff,\,1,2}$ which, respectively, separate the region with
expanding universes $(H>0)$ from the region with contracting universes
$(H<0)$ if $\rho_{m}\neq 0$.
The effective $\Lambda_{eff,1,2}$ becomes an asymptotic constant for
$t\rightarrow \infty$, if the conditions
\be 
 \label{3.7.21}
\frac{\dot{p}}{p}\longrightarrow\Sigma_{0}\,,\;\;\;\;
\frac{W(p)}{6p}\longrightarrow\Sigma_{1}\,,
\ee 
hold. From (\ref{3.7.20}), we get $\dot{H}\leq 0$ if
\be 
 \label{3.7.22}
H^{2}\geq -\frac{W(p)}{6p}\,.
\ee 
Conditions (\ref{3.7.21}) gives the asymptotic behaviour of field $p$
and potential $W(p)$.  By a little algebra, we  obtain that
asymptotically must be
\be 
 \label{3.7.24}
\Sigma_{0}=0\,,\;\;\;\;\;f(R)=f_{0}(R+6\Sigma_{1})\;;
\ee 
where $f_{0}$ is an arbitrary constant. The asymptotic solution is then
\be 
 \label{3.7.25}
H^2=\Sigma_{1}\,,\;\;\;\;\;p=p_{0}\,,\;\;\;\;\;\dot{H}=0\,.
\ee 
From Eq.(\ref{3.7.14}), 
or, equivalently, from the constraint (\ref{3.7.8}), we get
\be 
 \label{3.7.26}
R=-12H^2=-12\Sigma_{1}\,.
\ee 
The no--hair theorem is restored without using Bianchi 
identities (\ie the Klein--Gordon equation). The de Sitter
solution of the Einstein gravity is exactly recovered if
$\Sigma_{1}=\Lambda/3.$
 It depends on the free constant $f_{0}$ in (\ref{3.7.24}) which is assigned
by introducing ordinary matter in the theory. 
This means that, asymptotically, 
\be 
\label{desit}
f(R)=f_{0}(R+2\Lambda)\,.
\ee 
The situation is not completely
analogue to the scalar--tensor case~\cite{lambdat} 
since the request that asymptotically
$a(t)\rightarrow\exp(\Lambda t)$, univocally "fixes" the asymptotic 
form of $f(R)$. Inversely, any fourth--order theory which asymptotically
has de Sitter solutions, has to assume the form (\ref{desit}).
We have to stress the fact that it is
the {\it a priori} freedom in choosing $f(R)$ which allows to recover an
asymptotic cosmological constant (which is not  present in the
trivial case $f(R)=R$, unless it is put by hand) so that de Sitter
solution is, in some sense, intrinsic in higher--order theories.
A pure higher than fourth--order gravity theory is recovered, for example,
 with the choice
\be 
 \label{6.2}
{\cal A}=\int d^4x\sqrt{-g} f(R,\Box R)\,.
\ee 
which is, in general, an eighth--order theory. If $f$ depends only
linearly on $\Box R$, we have a sixth--order theory. 
As above, we can get a FRW pointlike Lagrangian with the position
${\l}={\l}(a,\dot{a},R,\dot{R},\Box R,\dot{(\Box R)}).$
Also here, we consider $R$ and $\Box R$ as two independent fields and use the
method of  Lagrange multipliers to eliminate higher derivatives than one 
in time.
We obtain the Helmholtz--like Lagrangian
\be 
\label{6.6}
{\l}=a^{3}\left[f-R{\g}+6H^2{\g}+6H\dot{\g}-\frac{6k}{a^2}{\g}-\Box R{\cal F}
-\dot{R}\dot{\cal F}\right]\,,
\ee 
where
${\displaystyle {\g}=\frac{\pa f}{\pa R}+\Box\frac{\pa f}{\pa \Box R}}$
and 
${\displaystyle {\fo}=\frac{\pa f}{\pa \Box R}}$.
The equations of motion are
\be 
 \label{6.10}
H^2+H\left(\frac{\dot{\g}}{\g}\right)+\frac{\chi}{6{\g}}=0\;;
\ee 
\be 
 \label{6.14}
\dot{H}=-\half\left(H^2+\frac{\chi}{6{\g}}\right)-
\half\left(\frac{\dot{\g}}{\g}\right)^{2}-
\frac{1}{2}\frac{d}{dt}\left(\frac{\dot{\g}}{\g}\right)-
-\frac{\dot{R}\dot{\fo}}{2{\g}}\;,
\ee 
\be 
 \label{6.8}
R=-6[\dot{H}+2H^2]\,,
\;\;\;\;
\Box R=\ddot{R}+3H\dot{R}\,,
\ee 
where Eqs.(\ref{6.8}) have the role of  Klein--Gordon
equations for the fields $R$ and $\Box R$ and are also "constraints" for
such fields.
The quantity $\chi$ is 
$\chi=R{\g}+{\fo}\Box R-\dot{R}\dot{\fo}-f.$ 
We can define an effective cosmological constant as 
\be 
\label{6.12}
\Lambda_{eff,\,1,2}=-\frac{\dot{\g}}{2{\g}}\pm
\sqrt{\left(\frac{\dot{\g}}{2{\g}}\right)^{2}-\frac{\chi}{6{\g}}}\,.
\ee 
$\Lambda_{eff,\,1,2}$ become asymptotically constants
if
\be 
 \label{6.13}
\frac{\dot{\g}}{\g}\longrightarrow\Sigma_{0}\,,\;\;\;\;
\frac{\chi}{6\g}\longrightarrow\Sigma_{1}\,.
\ee 
From (\ref{6.14}), we have $\dot{H}\leq 0$ if
\be 
 \label{6.15}
H^{2}\geq 
-\frac{\chi}{6{\g}}\,,\;\;\;\;\;\frac{\dot{R}\dot{\fo}}{\g}\geq 0\,.
\ee 
The quantities $\chi$, ${\g}$, and $\Lambda_{eff}$
are functions of two fields and the de Sitter asymptotic regime selects
particular surfaces $\{R,\Box R\}$.
In conclusion,
if we compare the conditions in ~\cite{wald} with ours, we have that
\bea
\mbox{(no--hair)}\;\;\;\;\;\; & &\;\;\;\mbox{(our asymptotic conditions)}
\nonumber\\
\;\;\left(H-\sqrt{\frac{\Lambda}{3}}\right)\left(H
+\sqrt{\frac{\Lambda}{3}}\right)\geq 0\;\;\;\;\;\;\; 
&\;\;\;\Longleftrightarrow\;\;\;&\;\;\;\;\;\;(H-{\Lambda}_{1})
(H-{\Lambda}_{2})\geq 0,\nonumber\\
\;\;\;\;\;\;\;\;\;\;\;\;\;\;\;\;\;\;\dot{H}\leq\frac{\Lambda}{3}-H^{2}\leq 0 
&\;\;\;\;\;\Longrightarrow\;\;\;&\;\;\;\;\;\;\; \dot{H}\leq 0\,.\nonumber
\eea
and in this sense the no--hair theorem is extended.
The above discussion can be realized in specific cosmological models.
 As in~\cite{lambdat} for scalar--tensor theories,  it is possible 
to give  examples where, by fixing the 
higher--order theory, the asymptotic de Sitter 
regime is restored in the framework  of no--hair
theorem.
The presence of standard fluid matter can be implemented by adding the
term
${\displaystyle {\cal L}_{m}=Da^{3(1-\gam)}}$
into the FRW--pointlike Lagrangian.
It is a sort of pressure term. 
For our purposes it is not particularly
relevant.
The conditions for the existence and stability of de Sitter solutions
for fourth--order theories $f(R)$ are widely discussed~\cite{ottewill}. 
In particular, it is shown that, for $R$ covariantly constant 
(\ie $R=R_{0}$),
as recovered in our case for $R\rightarrow-12\Sigma_{1}$,
the field equations  yield the existence condition 
$R_{0}f'(R_{0})=2f(R_{0})$.
Thus, given any $f(R)$ theory, if there exists a solution $R_{0}$, 
the theory contains a de Sitter solution.
From our point of view, any time that the ratio  $\dot{f}(R(t))/f(R(t))$ 
converges
to a constant, a de Sitter (asymptotic) solution exists.
On the other hand, given, for example, a theory of the form
${\displaystyle f(R)=\Sigma_{n=0}^{N}a_{n}R^{n}}$,
the above condition  is satisfied if the polynomial
equation
${\displaystyle \Sigma_{n=0}^{N}(2-n)a_{n}R_{0}^{n}=0}$,
has real solutions. Examples of de Sitter asymptotic behaviours
recovered in  this kind of theories are given in literature~\cite{coa}.
Examples of theories higher than fourth--order in which asymptotic de Sitter
solutions are recovered are discussed in~\cite{kluske}.
There is discussed under which circumstances the de Sitter space--time
is an attractor solution in the set of spatially flat FRW models.
Several results are found: for example, a $R^{2}$ non--vanishing term
is necessarily required (\ie a fourth--order term cannot be escaped);
the models are independent of dimensionality of the theory;
more than one inflationary phase can be recovered.
Reversing the argument from our point of view, a wide class of
cosmological models coming from higher--order theories, allows to recover
an asymptotic cosmological constant which seems an intrinsic feature
if Einstein--Hilbert gravitational action is modified by higher--order terms.
In this sense, and with the conditions given above, the cosmological
no--hair theorem is extended.

\vfill

\end{document}